\documentclass[aps,prb,twocolumn,showpacs,superscriptaddress]{revtex4-1}
\usepackage{amssymb,amsfonts,amsmath}
\usepackage{graphicx}
\usepackage{epstopdf}
\usepackage{tabularx}
\usepackage{bm}
\usepackage{euscript}
\usepackage{epsfig,psfrag,subfigure}
\usepackage{color}
\usepackage{amsfonts}
\usepackage{exscale}
\usepackage{amsbsy}
\usepackage{stmaryrd}
\usepackage{wasysym}
\usepackage{verbatim}
\usepackage{xcite}
\usepackage{xr}

\def\be{\begin{equation}}       \def\ee{\end{equation}}
\def\bea{\begin{eqnarray}}      \def\eea{\end{eqnarray}}
\def\ba{\begin{array} }
\def\ea{\end{array} }
\def\bnum{\begin{enumerate} }
\def\enum{\end{enumerate}}

\def\=>{\Rightarrow}
\def\>{\rightarrow}

\def\eye2{Fathbb{I}}

\begin{document}

\title{Self-Duality and a Hall-insulator phase near the superconductor-to-insulator transition in indium-oxide films}

\author{Nicholas P. Breznay}
\affiliation{Materials Science Division, Lawrence Berkeley National Laboratory, Berkeley, CA 94720.}
\author{Myles A. Steiner}
\affiliation{National Renewable Energy Laboratory, 15013 Denver West Parkway, Golden, CO 80401.}
\author{Steven A. Kivelson}
\affiliation{Department of Physics, Stanford University, Stanford, CA 94305, USA}
\author{Aharon Kapitulnik} 
\affiliation{Department of Physics, Stanford University, Stanford, CA 94305, USA} 
\affiliation{Department of Applied Physics, Stanford University, Stanford, CA 94305, USA}
\date{\today}

\begin{abstract} 
We combine measurements of the longitudinal ($\rho_{xx}$) and Hall ($\rho_{xy}$) resistivities  of disordered  two dimensional amorphous indium-oxide films   to study the magnetic-field tuned superconductor to insulator transition (H-SIT) in the $T \to 0$ limit.  At the critical field, $H_c$, the full resistivity tensor is $T$ independent with $\rho_{xx}(H_c) = h/4e^2$ and $\rho_{xy}(H_c)=0$ within experimental uncertainty in all films (i.e. these appear to be ``universal'' values); this is strongly suggestive that  there is a particle-vortex self-duality at $H=H_c$.  The transition separates the (presumably) superconducting state at $H<H_c$  from a ``Hall-insulator" phase in which $\rho_{xx}\to \infty$ as $T\to 0$ while $\rho_{xy}$ approaches a non-zero value smaller than its  ``classical value" $H/nec$, i.e. $0<\rho_{xy}<H/nec$.  A still higher characteristic magnetic field, $H_c^*> H_c$,  at which the Hall resistance  is $T$ independent and roughly equal to its classical value, $\rho_{xy}\approx H/nec$,  marks an additional crossover to a highfield regime (probably to a Fermi-insulator) in which $\rho_{xy} > H/nec$ and possibly diverges as $T\to 0$.  We also highlight a profound  analogy between the H-SIT and quantum-Hall liquid to insulator transitions (QHIT).
\end{abstract}

\maketitle

\section{Significance Statement}
{\it The magnetic-field tuned superconductor-to-insulator transition (H-SIT) is a paradigmatic quantum phase transition and, along with the quantum-Hall liquid-to-insulator transitions (QHIT), is among the best experimentally studied ones. However, in the transition and the proximate ground-state phases, it has consistently exhibited features that are seemingly at odds with the generally accepted theoretical ``story." The clear evidence we have found of particle-vortex duality at the H-SIT is one such example, as is the associated evidence that the proximate insulating phase is fundamentally distinct from a conventional ``Anderson insulator'' in that $\rho_{xy}$, rather than diverging, tends to a finite value as T$\rightarrow$0. That these features are analogous to behaviors previously documented near the QHIT supports the existence of the correspondence between the two problems implied by the composite boson theory.}

\section{Overview}
Quantum phase transitions (QPTs) occur at zero temperature ($T=0$) as a quantum control-parameter  is varied. Where the transition is continuous, quantum critical phenomena are expected to give rise to universal physics which can be analyzed using a straightforward scaling theory. The magnetic field tuned transition between superconducting and insulating ground states in two-dimensional conductors is a particularly attractive exemplar of a QPT since the magnetic field can be continuously tuned, allowing a detailed scaling analysis of the QPTs and explorations of the ground state phases proximate to criticality \cite{Hebard1990,Gantmakher1998,Sambandamurthy2004,Steiner2005,Yazdani1995,Markovic1998,Bielejec2002,Baturina2004,Steiner2008}. However, the exact nature of the insulating and superconducting states above and below the H-SIT, and a satisfactory description of the transition between them are still lacking.

The conventional picture of $T\to 0$ phases of a two-dimensional electron fluid in the presence of disorder  is based on the assumption that the only stable phases are superconducting or insulating (or, in a magnetic field, quantum Hall liquid phases). In contrast, studies of films near the H-SIT have suggested the existence of several unexpected new ground-state phases in films that superconduct at zero field.  In weakly disordered films (with normal state resistivity small compared to the quantum of resistance, $\rho_N \ll h/e^2$), the superconducting state gives way to an ``anomalous metallic phase'' with a resistivity that  extrapolates to a non-zero value, $0<  \rho(T\to 0,H)\ll \rho_N$  \cite{Ephron1996,Mason1999,Eley2012,Liu2013,Han2014}. For  highly disordered superconducting films with $\rho_N \sim h/e^2$, a direct H-SIT seemingly occurs at a field $H_{c}$, but as we will discuss, significant electron ``pairing'' persists in the insulating phase. 

In a purely bosonic description~\cite{Fisher1990a,Lee1989,Wen1990} (where it is assumed that fermionic excitations are negligible),  the insulating state is characterized as a condensate of delocalized vortices and localized Cooper pairs, while the superconducting state is a condensate of Cooper pairs with localized vortices. Quantum fluctuations of the phase of the superconducting order parameter control this QPT. A key  feature of particle-vortex duality  is that the  (measured) conductivity tensor $\underline{\sigma}$ is equal to the vortex-resistivity tensor $\underline{\rho}^v$, 
\be
\underline{\sigma}=(4e^2/h)^2 \underline{\rho}^v.
\ee
An emergent {\em self-duality}  in the neighborhood of $H_c$ would imply $\underline{\sigma}^T(H_c+\Delta H)=\underline{\sigma}^v(H_c-\Delta H)$ ( $\underline{\sigma}^T$ is the transpose), or in other words at criticality $[\sigma_{xx}(H_c)]^2+[\sigma_{xy}(H_c)]^2= (4e^2/h)^2$.  If we further imagine that $\sigma_{xy}$ is continuous at $H=H_c$ it would follow that $\sigma_{xy}(H_c)=0$ since $\sigma_{xy}\to 0$ as $T\to 0$ in the insulating phase. (Analogous reasoning was used to infer  the critical conductivity tensors at the QHIT -- see
Supporting Information.)  Together, these arguments  imply
\bea
\rho_{xx}(H_c)={h/4e^2}  \ \ \ {\rm and} \ \ \ \rho_{xy}(H_c)=0
\label{dual}
\eea
Previous studies have examined evidence for duality \cite{Paalanen1992,Ovadia2013} from resistivity measurements, but have not examined the full conductivity tensor across the transition.

\subsection{Summary of Results} In this paper we provide  insights concerning the nature of the H-SIT in highly disordered films using new measurements of the full resistivity tensor across the quantum transition. (See Fig.~\ref{rxxrxy}.)  We draw three key conclusions: (1) We identify  the insulating state above $H_c$ as a ``Hall insulator''  \cite{Kivelson1992} in which $\rho_{xx}\to \infty$ as $T\to 0$, but $0<\rho_{xy}\leq H/nec$, cementing a connection  between the H-SIT and the QHIT. (2) We observe self-duality consistent with Eq.~\ref{dual} at the transition. (3) We present suggestive evidence that the  superconducting state is a ``vortex insulator'' (dual to the Hall insulator) in which $\rho_{xx} \to 0$ but $\sigma_{xy}$ approaches a finite value as $T\to 0$.

\subsection{Concerning the $T\to 0$ limit}
A general issue in  studies of ground-state phases and quantum critical phenomena is that  experiments are carried out at non-zero $T$, so all  results must be extrapolated to $T=0$.  There are numerous practical issues that define the lowest temperatures at which experiments can be carried out -- in addition to  issues of refrigeration, equilibration times (especially in disordered systems) tend to diverge rapidly with decreasing $T$ and the range of current densities for which linear response theory applies decreases.  In the present case, the fact that $ \rho_{xy}/\rho_{xx} \to 0$ as $T\to 0$ ultimately limits our ability to reliably measure $\rho_{xy}$, although using the distinct symmetries of $\rho_{xy}$ and $\rho_{xx}$ with respect to $H\to -H$ helps greatly in this regard.  Here, we report results at high enough temperatures that we avoid measurement ambiguities, and yet reach temperatures low compared to ``microscopic'' scales (for instance, low compared to the zero field $T_c$) so that we are well within the quantum critical fan that describes the basin of influence of the quantum critical point such that it is reasonable to extrapolate the results  to $T=0$.  We note that while this argument is compelling at criticality,  there is always an emergent energy scale which vanishes upon approach to criticality, so inferring the asymptotic properties of the stable phases near criticality is intrinsically subtle \cite{Sachdev2011}. 

\subsection{Critical Behavior and Scaling} 
\begin{figure}[ht]
\centering
\includegraphics[width=1.0\columnwidth]{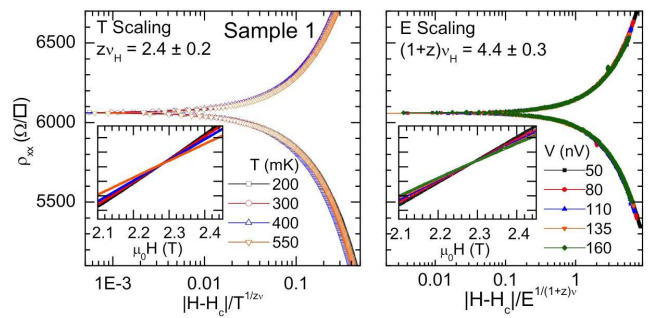}
\caption{ (Color online) Scaling of isotherms (left) and constant electric field curves (right) near the H-SIT for sample 1; temperature and applied bias voltages are indicated. Insets show the raw resistivity isotherm and constant electric field data, with the same vertical scale as in the main panel.}
\label{scaling}
\end{figure}

\ 
Near the H-SIT, it is reasonable to expect that the singular part of various physical quantities, most especially the diagonal resistivity, are well described by a universal scaling function with appropriate universal critical exponents.  Thus, in common with earlier studies of the H-SIT in highly disordered films,  we perform a scaling analysis of the $T$ and $H$ dependence of $\rho_{xx}$   in a narrow neighborhood of the critical resistivity (shown in Fig.~\ref{scaling}):
 \be
 \rho_{xx}(T,H)=\rho_c\ \mathcal{F}(X) \ {\rm where} \ X=(H-H_c)/T^{1/z\nu_H}.
 \label{F}
 \ee
 This yields  the combination of critical exponents $\nu_H z\approx 2.4$ (see Fig.~\ref{scaling}) and  a value of $\rho_c$ that is universal  (within experimental uncertainty) and equal to the ``Cooper-pair quantum of resistance,'' $\rho_c = h/(2e)^2\approx 6.45$ k$\Omega$ \cite{Rhoc}.  Moreover, from an additional scaling ansatz\cite{Yazdani1995} for the non-linear field dependence of the differential resistivity at criticality \cite{ep,Ovadyahu1982}, 
\be
\rho_{xx}=\rho_c\ \mathcal{G}(Y) \ {\rm  where} \ Y=(H-H_c)/E^{1/(1+z)\nu_H}, 
\label{G}
\ee
one can extract another combination of critical exponents,  $(z+1)\nu_H = 4.4\pm 0.3 $. Together, these results imply that the correlation length exponent $\nu \approx 2.3$ and the dynamical exponent $z\approx 1$.

Analogous behavior has been observed \cite{Wong1995,Sondhi1997} at various quantum Hall to insulator transitions (QHIT) in 2DEG systems.  In fact, a formal mapping between the two problems yields an analogy between the H-SIT and the QHIT \cite{Kivelson1992}.  As has been emphasized previously\cite{Kivelson1992}, it is striking that a scaling collapse of data from both the integer (filling factor 2) and the fractional ( filling factor 1/3) QHIT produce scaling curves that look extremely similar to those from the H-SIT in highly disordered films with the same value of $z\nu_H \approx 2.3$, with a different (but analogous) universal critical resistance $\rho_c = h/e^2$ \cite{QuantumPercolation}.  Significantly, the insulating phase proximate to the QHIT transition is an unconventional ``Hall insulator'' in the sense that while $\rho_{xx}\to \infty$ as $T\to 0$, $\rho_{xy}$ approaches a finite field-dependent value.\cite{Johnson1993,Shahar1997} 

\section{Experimental Results}
\begin{figure}[ht]
\centering
\includegraphics[width=1.0\columnwidth]{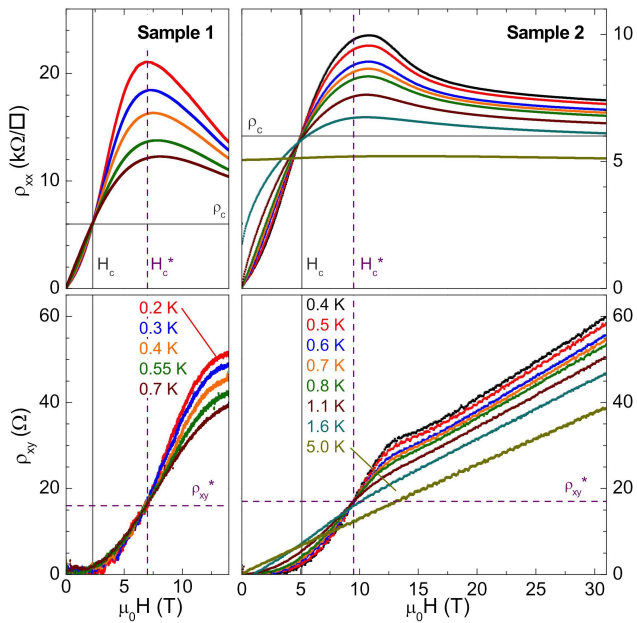}
\caption{ (Color online) Longitudinal ($\rho_{xx}$) and Hall ($\rho_{xy}$) resistances for two InO$_x$ samples. Solid lines mark the H-SIT field $H_c$ and critical resistivity $\rho_c$, and dashed lines mark the Hall-crossing field $H_c^*$ and resistivity $\rho_{xy}^*$.}
\label{rxxrxy}
\end{figure}

The amorphous InO$_x$ films used in the present study were synthesized and measured as described previously\cite{Steiner2005,Steiner2008} and further discussed in the Supporting Information. Figure~\ref{rxxrxy} depicts the basic structure of $\rho_{xx}(T,H)$ and $\rho_{xy}(T,H)$ for two representative InO$_x$ samples (1 and 2). (Comparable results were found on six additional samples measured during this study, and are similar to data previously published in Ref.~\cite{Paalanen1992}.)  In comparison with some other strongly disordered films (e.g. see \cite{Sambandamurthy2004,Steiner2005,Baturina2007b,Ovadia2015},)  the  magnetoresistance peaks  exhibited at accessible temperatures by the present films are only moderately large, which allows for accurate measurements of $\rho_{xy}$. (All $\rho_{xy}$ data are field-antisymmetric.) Strongly disordered samples with $\rho_N \sim  h/e^2$, which includes the two presented here, all show critical resistance at the H-SIT  within 10$\%$ of the quantum of resistance, $h/4e^2$ \cite{Rhoc}, and exhibit good scaling of the form of Eqs. \ref{F} and \ref{G}.
There are several important features in the data of Fig.~\ref{rxxrxy}. First, in addition to the hallmark crossing point of $\rho_{xx}$ at ($H_c$,$\rho_{c}$) marking the SIT, we observe at higher fields a crossing point of $\rho_{xy}$  at ($H_c^*,\rho_{xy}^*$). $H_c^*$ roughly coincides with the field at which the longitudinal magnetoresistance peaks, suggesting that it is associated with a {\it crossover} from Bose-dominated to Fermi-dominated behavior. While for $H_c<H<H_c^*$ \  $\rho_{xy}$ decreases with decreasing $T$, this dependence weakens as $T\to 0$, suggesting that $\rho_{xy}$ approaches a finite value. 

\subsection{Results extrapolated to $T \to 0$} \
In order to obtain a more explicit understanding of the nature of the different regimes above and below the H-SIT, and what can be inferred about the $T\to 0$ limit, we analyzed the full set of data that determines the resistivity tensor.
\begin{figure}[ht]
\centering
\includegraphics[width=1.0\columnwidth]{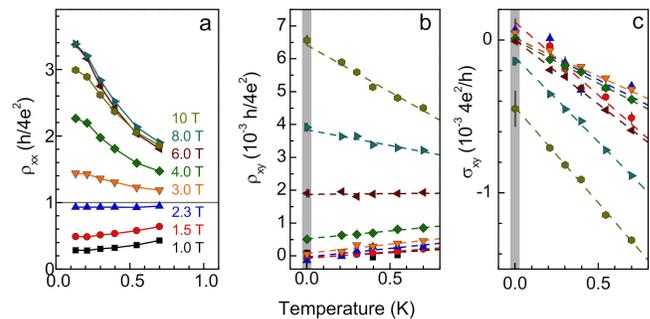}
\caption{ (Color online) $T$ dependence of the resistivity tensor ( $\rho_{xx}$ in a) and $\rho_{xy}$ in b) and $\sigma_{xy}$ (in c)  of Sample 1 in units of the superconducting quanta for various values of $H$.  For this sample, $H_c = 2.3$ T, $H_c^*= 7.0$ T, $T_c(H=0)=1.1$K, and $\rho_N=0.27 h/4e^2$.}
\label{extrapolate}
\end{figure}

\begin{figure}[ht]
\centering
\includegraphics[width=1.0\columnwidth]{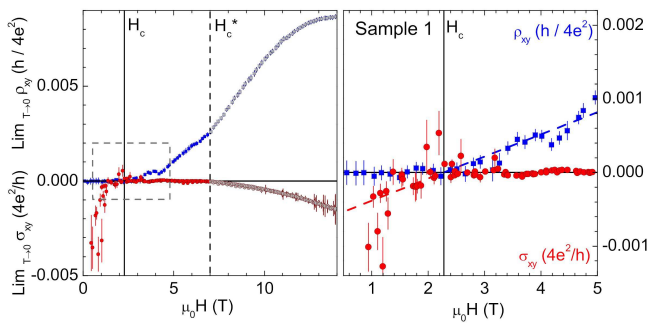}
\caption{ (Color online) The  $T \to 0$ extrapolated values of $\rho_{xy}(H)$ and $\sigma_{xy}(H)$ for sample 1.The vertical solid lines mark the SIT, and dashed lines the Hall crossing field. The right panel highlights the SIT region (marked in the left panel);  lines through the data are guides to the eye, showing the symmetry between $\rho_{xy}$ and $\sigma_{xy}$. }
\label{duality}
\end{figure}

In Fig.~\ref{extrapolate} we present the $T$ dependence of the resistivity tensor of sample 1 at various fixed values of $H$. Note that, having measured both, $\rho_{xx}$ and $\rho_{xy}$, we can calculate the off-diagonal term of the conductivity tensor: $\sigma_{xy} = -\rho_{xy}/\left(\rho_{xx}^2 + \rho_{xy}^2\right)$.  Fig.~\ref{extrapolate} shows $\rho_{xx}$, $\rho_{xy}$, and $\sigma_{xy}$ of sample 1 for various fixed fields as a function of $T$, as well as sketching ways in which we infer $T\to 0$ values  by extrapolation. We now distinguish several field regimes describing the different behavior of $\rho_{xx}$ and $\rho_{xy}$ as the magnetic field is increasing from low fields to much above $H_c^*$.
 Figure~\ref{duality} summarizes the groundstate Hall response as a function of $H$ based on a linear extrapolation of the data to $T=0$ (i.e. according to the  dashed lines in Fig.~\ref{extrapolate}).  Error bars reflect statistical uncertainty in the extrapolation procedure. On the high field side of the H-SIT ($H_c<H<H_c^*$), where presumably $\sigma_{xx} \to 0$ and $\rho_{xx} \to \infty$,  $\rho_{xy}$ approaches a finite limit which is greater than 0 and less than its classical value $H/nec$, while $\sigma_{xy} \to 0$  - these are the defining features of a ``Hall Insulator" phase \cite{Kivelson1992} (The value of the Hall resistance is taken from measurements at $T>T_c(H=0)$;  for sample 1, $1/nec= 1.2 \Omega/T$, i.e. $H_c^*/nec = 8.2 \Omega$ for $H_c^*= 7 T$.)  At the low field edge of the Hall insulating regime, $\rho_{xy}(H)\to 0$  as $H\to H_c^+$, and it grows monotonically with increasing $H$, approaching roughly its classical value at the high field boundary of the regime.

Conversely, in the low-field phase ($H<H_c$), $\rho_{xy} \to 0$, while $\sigma_{xy}$ approaches a finite limit which tends to 0 as $H\to H_c^-$. If we accept the identification of this phase as superconducting (although this is not entirely established in the accessible range of $T$), then we expect  that $\rho_{xx} \to 0$ and $\sigma_{xx}\to \infty$ as $T\to 0$, i.e. that the conductivity tensor on the low field side of the H-SIT behaves analogously to the resistivity tensor on the high field side \cite{Fisher1990b,Schakel2000}. 
 
 There also appears to be a second critical (or crossover) field $H_c^*> H_{c}$ at which a transition to a ``Fermionic insulator'' occurs.  For $H>H_c^*$, $\rho_{xy}$ (as well as $\rho_{xx}$)  grows with decreasing $T$.  While the $T$ dependence of  $\rho_{xx}$ clearly suggests that it diverges as $T\to 0$, the much weaker $T$ dependence of $\rho_{xy}$, and our limited temperature range, make it less clear whether for $H>H_c^*$ it diverges. In any case, the upward curvature of $\rho_{xy}(T)$ suggests that the value  shown in Fig. \ref{duality}, obtained from linear extrapolation to $T=0$, underestimates the true value.

We now look with more detail at the $T$ dependence of the resistivity tensor of sample 1 in the various distinct field ranges, particularly those aspects that affect the extrapolation $T\to 0$:

\subsection{$0<H<H_c$} \ Here $\rho_{xx}$ is a  decreasing function of decreasing $T$.  However, the $T$ dependence of $\rho_{xx}$ is sufficiently weak that a linear extrapolation to $T=0$ would imply a non-zero magnitude of $\rho_{xx}(T=0)$, which would be inconsistent with our identification of this as a superconducting phase.  This issue appears to plague all measurements of strongly disordered films\cite{Sambandamurthy2004,Steiner2005,Baturina2007a}, although it has not been commented on previously.  (We will return to this point in the discussion, below.)  $\rho_{xy}$ in this range of $H$ is always very much smaller than  its classical value ($\rho_{xy} \ll H/nec$) and  clearly tends to zero as $T\to0$. Inverting the resistivity tensor to obtain $\sigma_{xy}$ amplifies the error bars.  Of course, since $\rho_{xy} > 0$, $\sigma_{xy}$ is negative;  it is also small and approximately a linear function of $T$, but a linear extrapolation of $\sigma_{xy}$ to $T=0$ (the dashed lines in Fig.~\ref{extrapolate}) suggests that  $\sigma_{xy} $ approaches non-zero negative values, as shown in Fig.~\ref{duality}.

\subsection{$H\approx H_c$} \  At this magnetic field $\rho_{xx}(H_c,T)$ is essentially $T$ independent and approximately 10\% less than $h/4e^2$.  As mentioned above, the sign and magnitude of  $[\rho_c-h/4e^2]$ varies from sample to sample, but is rarely larger than this. While the resistance can be measured with much better than 10\% accuracy, extracting the resistivity requires precise knowledge of the geometry of the current pathways.   In the present experiments,  this geometric uncertainty is at least equal to the apparent deviations from universality.  Both $\rho_{xy}$ and $\sigma_{xy}$ are vanishingly small within experimental accuracy, i.e. $|\rho_{xy}(H_c)/(H_c/nec)| \ll 1$ and $|\sigma_{xy}(H_c)\cdot(H_c/nec)| \ll 1$.

\subsection{$H_c < H<H_c^*$} \  In this regime $\rho_{xx}$ is a strongly increasing function of decreasing $T$, with pronounced upward curvature, and low $T$ magnitudes that are large compared to the quantum of resistance. This behavior  identifies this as an insulating state, with a $T$ dependence that is consistent with  activated behavior or various forms of variable range hopping, $\rho_{xx} \sim \exp[(T_0/T)^\delta]$ with $\delta = 1$, 1/2, or 1/3 \cite{Steiner2005}.  However, the data are not consistent with any reasonable power-law and we do not find the super-Arrhenius behavior that has been reported in some systems \cite{Vinokur2008,Baturina2013}, suggesting that we are indeed probing an equilibrium phase.  $T_0$ as a function of $H$ grows continuously from $T_0=0$ at $H=H_c$ to a maximal value at around $H=H_c^*$.   If we adopt $\delta=1/3$, appropriate for Mott variable-range-hopping in 2D, we find for sample 1 that $T_0(H_c^*)= 0.5\pm0.1$K, comparable to the zero field transition temperature, $T_c$. This is highly suggestive that superconducting pairing remains significant even in the Hall insulating regime.  $\rho_{xy}$ is a weakly decreasing function of decreasing $T$, with a magnitude that is always less than its classical value, $H/nec$. Indeed, $\rho_{xy}$ is, within experimental uncertainty, a linear function of $T$, which extrapolates to the finite zero temperature value, $\rho_{xy}(T=0,H)$, shown in Fig. \ref{duality}, which grows monotonically with $H$ from 0 at $H=H_c$ to its classical value at $H=H_c^*$.  By contrast, $\sigma_{xy}$ extrapolates roughly linearly to values indistinguishable from 0 as $T\to 0$.

\subsection{$H=H_c^* > H_c$}  \ This is the field at which $\rho_{xy}$ is approximately $T$ independent, reflecting the crossover from a low field regime where $\rho_{xy}$ decreases with decreasing $T$ to a high field regime where it increases.  It is also roughly the value of $H$ at which $\rho_{xx}$ achieves its largest value for fixed $T$.  This large magnitude insures that $\sigma_{xy} \to 0$ as $T\to 0$.

\subsection{$H > H_c^*$} \ Here $\rho_{xy}$ is an increasing function of decreasing temperature, with a magnitude that is larger than its classical value. Over the accessible range of $T$, it can be roughly fit to a linear function, which results in the $T\to 0$ extrapolated values shown in Fig.~\ref{duality}.  However, the clear upward curvature likely indicate that this represents an underestimate, and it is even plausible that in this entire range, $\rho_{xy} \to \infty$ as $T\to 0$.  $\rho_{xx}$ is also an increasing function of decreasing $T$, but at fixed $T$ it is a decreasing function of increasing $H$, i.e. the film is increasingly metallic at higher fields.   A linear extrapolation of $\sigma_{xy}$ to $T=0$ would imply a non-zero value, as shown in Fig.~\ref{duality}.  At present, it is not clear what to conclude about the nature of the ground-state behavior in this regime.

Indeed, in Fig.~\ref{rxxrxy} data is presented on Sample 2 up to 32T, i.e. to fields much higher than any estimate of a mean-field $H_{c2}$.  Here, one can see that $\rho_{xx}$ takes on values that are significantly smaller than the electron resistivity quantum, $h/e^2 = 25.8$k$\Omega$, yet much larger than the normal state value, $\rho_N \approx 5 $k$\Omega$.  None-the-less, $\rho_{xx}$ shows an ``insulating-like'' $T$ dependence.  Moreover, while $\rho_{xy}$ is a linearly increasing function of $H$ with an almost $T$ independent slope (plausibly giving a measure of $1/nec$), it has a peculiar extrapolated $H\to 0$ offset which grows with decreasing $T$.  It is not at all clear what the nature of the state is that gives rise to these behaviors.

\subsection{Global consistency check} \ 
The success of the scaling analysis near criticality supports the assertion that the accessible range of $T$ is sufficiently low to penetrate well into the quantum critical regime.  However, the relatively weak $T$ dependence of $\rho_{xx}$ in the putative ``superconducting'' regime ($H<H_c$), or similarly weak $T$ dependence of $\rho_{xx}$ on the insulating side of the transition ($H_c < H<H_c^*$), may suggest that the temperatures probed are not yet sufficiently low to fully sense the character of the respective ground-states. Thus, we introduce a simple ansatz for the $T$ and $H$ dependence of $\underline{\rho}$, which presupposes the existence of a
 $T=0$ H-SIT  with particle-vortex duality  to test the self-consistently of this assumption.

\begin{figure}[ht]
\centering
\includegraphics[width=1.0\columnwidth]{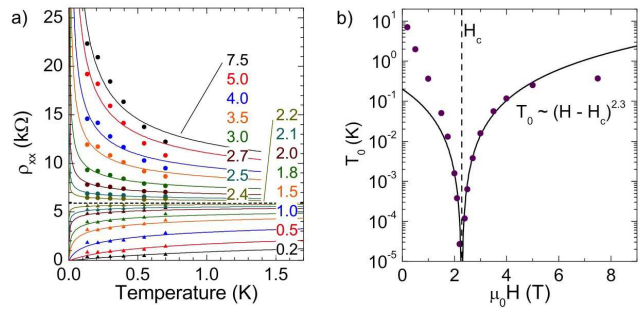}
\caption{ (Color online) (a) Fits of the low temperature scaling of $\rho_{xx}(T)$ curves using equations \ref{ansatz}  and \ref{ansatz2}. (b) The characteristic temperatures extracted from the fits to equations \ref{ansatz} and \ref{ansatz2} as a function of magnetic field, showing a critical behavior at $H_c$ consistent with the critical exponent $\nu_H z \approx 2.3$ found in Fig.~\ref{scaling}.}
\label{scalingb}
\end{figure}

Starting on the insulating side of the transition, we consider it most likely that the resistance is dominated by variable-range-hopping of Cooper pairs:
\be
\rho_{xx}(T,H) \approx \rho_c\exp[(T_0(H)/T)^{\delta}] \ {\rm for} \ H>H_c
\label{ansatz}
\ee
with $\delta=1/3$, and we assume that 
\be
\rho_{xy}(T,H) \approx \rho_{xy}(0,H) + {\cal O}(T) \ {\rm for} \ H>H_c
\ee
(which of course implies that $\sigma_{xy}\to 0$ as $T\to 0$). We can already see from Fig.~\ref{extrapolate} that this ansatz gives a good account of the $H$ and $T$ dependence of $\rho_{xy}$ in this range of fields, and indeed from Fig.~\ref{duality} it is clear that near the H-SIT, $\rho_{xy}(0,H) \approx 6.5\times 10^{-5}\rho_c[H-H_c]/H_c$.  In Fig.~\ref{scalingb}a we exhibit the quality of the fit obtained setting $\delta=1/3$, and treating $T_0(H)$ as a fitting parameter.  While there are  differences between the results based on this ansatz and the data. especially at higher $T$,  given that variable-range-hopping is a low $T$ asymptotic, and the (excessive) simplicity of the ansatz invovled, the fit is quite acceptable.  Notice in the second panel (Fig.~\ref{scalingb}b), the dependence of $T_0$ on $H$ is consistent with scaling close to $H_c$, i.e. $T_0(H) \sim [H-H_c]^{\nu z}$ with $\nu z\approx 2.3$.  This confirms that the data are at least consistent with the existence of an insulating phase as $T\to 0$ for $H>H_c$.

Moving to the superconducting side of the transition, we invoke duality to describe the resistivity tensor in terms of variable-range-hopping of vortices.  In other words, we introduce the ansatz
\be
\sigma_{xx}(T,H) \approx \rho_c\exp[(T_0(H)/T)^{\delta}] \ {\rm for} \ H<H_c
\label{ansatz2}
\ee
again with $\delta=1/3$, and we assume that 
\be
\sigma_{xy}(T,H) \approx  -\sigma_{xy}(0,H) + {\cal O}(T) \ {\rm for} \ H<H_c.
\ee
and by implication, $\rho_{xy}\to 0$ as $T\to 0$.  The consistency of this ansatz with the Hall data can again be read off of Figs.~\ref{extrapolate} and \ref{duality}, and the comparison for $\rho_{xy}$ is shown in Fig.~\ref{scalingb}.   The apparent good quality of the fit reinforces the assumption that the low field phase is superconducting.  Notice that away from criticality, the accessible temperatures extend well below $T_0(H)$;    the experimentally observed weak $T$ dependence stems from the small exponent, $\delta=1/3$, rather than from being at a larger $T$ than characterizes the superconducting state.

\section{Discussion}
Quantum fluctuations of the superconducting order parameter ultimately drive the transition from the superconducting state; the long-distance properties of these fluctuations are described by a complex (bosonic) scalar field  which loosely speaking represents the Cooper pairs.  A dual description of the same degrees of freedom can be given in terms of vortex variables \cite{Fisher1990b,Lee1989,Wen1990}.  In the superconducting phase, the vortices are localized and the Cooper pairs are condensed, while in an insulating phase, the vortices are condensed and the Cooper pairs are localized \cite{Fisher1990a}.  A similar situation pertains to the QHIT in the context of the composite boson formulation of the problem:  here, the quantized Hall plateau phase in which $\sigma_{xx}=0$ and $\sigma_{xy}=e^2\nu/h$ with $\nu = 1$ (or $1/3$ or etc.) corresponds to the condensed phase of the appropriate form of composite bosons.  Deep in this phase, isolated vortices are identified as  (possibly fractionally charged)  localized quasi-holes.   Thus, the QHIT is equivalent to a SIT transition of composite bosons, albeit with the  difference that the composite bosons are coupled to an emergent Chern-Simons gauge field. 
The observation that the conductivity-resistivity duality relation is satisfied within remarkably tight error bars wherever a direct SIT is observed (as demonstrated  in Fig.~\ref{duality}) and that the analogous relations are satisfied at the  transition points in a number of quantum Hall experiments is strong evidence that the critical theory is self-dual.

We  have interpretted our results as reflecting primarily collective order parameter (Cooper pair) fluctuations, neglecting the role of gapless quasiparticles. This interpretation is  plausible, given that (for all the samples used in this study), $H_c$ is much smaller than the estimated mean-field critical field, $ H_{c2}$.\cite{Sambandamurthy2004,Steiner2005} 
The strong positive magnetoresistance of the insulating phase close to $H_c$ is also highly suggestive that substantial pairing persists for a wide range of fields on either side of $H_c$.  Thus it is plausible that gapless quasiparticle  degrees of freedom do not play a significant role in the quantum dynamics in the neighborhood of the SIT.  By contrast, such quasiparticles are thought to play a key role in the anomalous metallic phase in weakly disordered films \cite{Mason1999,Kapitulnik2001,Tewari2005}.  Applying the analysis of Ref. \cite{Kivelson1992} for the QHIT, we note that if  the  quantum  critical point is self-dual, and both, $\rho_{xy}$ and $\sigma_{xy}$ are continuous functions of magnetic field, then the insulating phase proximate to the SIT will exhibit a finite $\rho_{xy}$ that rises continuously as a function of increasing $H$ for  $H > H_c$, while the superconducting phase will exhibit a finite $\sigma_{xy}$ that increases continuously with decreasing $H$ for  $H < H_c$.   

There have been several attempts to derive the properties of the Hall Insulator directly.  It was shown \cite{Zhang1992} for a  an Anderson insulator,  that in the non-canonical order of limits, first $T\to 0$ and then $\omega\to 0$, that $\rho_{xy}\sim H/nec$.  However,  in experiments, the resistivity is measured in the zero frequency limit at finite $T$ and then the results are extrapolated to the $T\to 0$ limit.  It has been shown that for the conventional theory of variable range hopping $\rho_{xy}\to \infty$ as $T\to 0$, although this divergence is much slower than the divergence of $\rho_{xx}$ \cite{Entin1995}.  This suggests the possibility the ``break" in the Hall resistance at $\rho_{xy}(H_c^*)$ marks  the transition  from a bosonic Hall insulator to a more conventional Anderson insulator.
Conversely, an analysis of vortex dynamics in a weakly superconducting state by Vinokur {\it et al.} \cite{Vinokur1993} lead to the conclusion that it gives rise to a non-vanishing value of $\sigma_{xy}$ as $T\to 0$; as these authors already pointed out, duality maps this behavior for $H<H_c$ to Hall insulating behavior for $H>H_c$.  

\section{Phase Diagram}
\begin{figure}[ht]
\centering
\includegraphics[width=1.0\columnwidth]{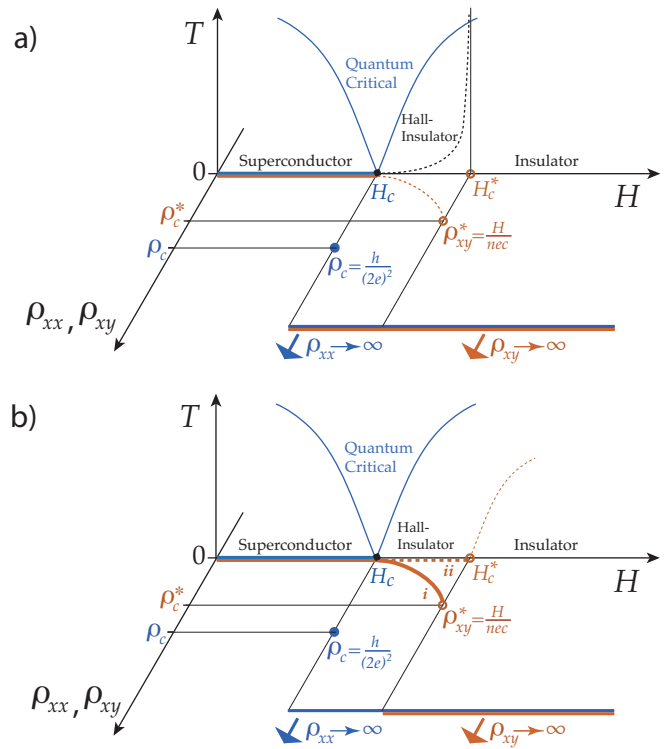}
\caption{ (Color online) Possible (T,H) phase diagrams for the SIT based on the $T\to 0$ trends of $\rho_{xx}$ and $\rho_{xy}$;  (a) A scenario where superconductivity is lost at $H_c$ due to phase fluctuations. A Hall-insulating phase appears near the SIT, but it crosses over to a true insulating phase at high fields. (b) A true Hall insulating phase exists in the field  range $H_c < H<H_c^*$. Here we show the possibility of a ``standard'' Hall insulator phase in ``{\it i}'' and a ``quantized'' Hall insulating phase in ``{\it ii}''.}
\label{qcc}
\end{figure}

The most straightforward scenario is depicted in Fig.~\ref{qcc}a; here superconductivity is lost at $H_c$ due to phase fluctuations, but the amplitude fluctuates slowly enough that we can consider Cooper pairs as still maintaining their identity, similar to a Kosterlitz-Thouless transition \cite{Fisher1990a}. In this case, a Hall-insulating phase appears near the SIT, but it crosses over to a true insulating phase, below a low crossover temperature (shown as the dashed line in the figure), below which $\rho_{xy}$ would begin to grow, making this phase ultimately no different from the fermionic insulating phase expected at higher fields. 

A different scenario is depicted in Fig.~\ref{qcc}b in which the Hall insulator for $H_c < H<H_c^*$ is taken to be a distinct phase, characterized by a finite zero T value of $\rho_{xy}$.  In this case we distinguish a ``standard" Hall insulator phase \cite{Kivelson1992} (indicated with the solid line, labeled ``{\it i}"), and a ``quantized Hall insulator" phase  (indicated with the dashed line, labeled ``{\it ii}"). Since there is always uncertainty in extrapolation $T\to 0$ it is still possible that $\rho_{xy}\to 0$ for $H_c < H<H_c^*$, in which case this phase could be classified as ``quantized" \cite {Shimshoni1997,Hilke1998} with $\rho_{xy}\equiv 0$. For $H_c^*<H$, both $\rho_{xx}$ and $\rho_{xy}$ increase (likely diverging) as $T \to 0$ as expected for a fermion-dominated insulator.

\section{Conclusions}
We have shown that when the Hall effect can be measured near the H-SIT:  1) the resistivity tensor at criticality approaches the universal value expected at a point of vortex-particle self-duality, 2) the critical exponents $\nu_H$ and $z$ appear to be the same as those observed at both the integer and fractional QHIT, and 3) the insulating phase proximate to the SIT appears to be a Hall insulator in which $\rho_{xx} \to \infty$ and $\rho_{xy}$ is finite as $T\to 0$, approaching $\sim H/nec$ with increasing field.

Finally, we observe that our data are consistent with the existence of a second quantum phase transition at $H=H_c^*$. This would give a natural explanation for the sharp change in behavior of $\rho_{xy}$ and imply that the Hall insulator should be taken to be a distinct quantum phase of matter. However, we cannot rule out  the possibility that $H_c^*$ is the point of a crossover at which unpaired electrons reassert their significance~\cite{maybe}.

We acknowledge important discussions with Boris Spivak, Yigal Meir, Dan Shahar, and especially important input from Mike Mulligan, and experimental assistance from Alexey Suslov at the NHMFL DC Field facility. Initial work was supported by the NSF. This work was supported by the Department of Energy Grant DE-AC02-76SF00515. A portion of this work was performed at the National High Magnetic Field Laboratory, which is supported by the National Science Foundation Cooperative Agreement No. DMR-1157490, the State of Florida, and the U.S. Department of Energy. 

\clearpage

\renewcommand{\thesection}{\Alph{section}}
\renewcommand{\thesubsection}{\thesection\arabic{section}}
\renewcommand{\thesubsubsection}{\thesection\arabic{subsection}.\alph{subsubsection}}
\renewcommand\thefigure{S\arabic{figure}}
\renewcommand\theequation{S\arabic{equation}}
\setcounter{section}{0}
\setcounter{figure}{0}
\setcounter{equation}{0}

\begin{center}
\large{Supporting Information\\
{\normalsize for}\\
\textbf{Self-Duality and a Hall-insulator phase near the superconductor-to-insulator transition in indium-oxide films}}
\end{center}

\section{EXPERIMENTAL}
\subsection{Materials and Methods}

Films were prepared by electron beam evaporation of sintered InOx onto an acid-cleaned silicon-nitride substrates. Control of the amount of disorder (hence, the ``strength" of the SIT) is achieved by adding oxygen during growth, and then subsequent careful, low-temperature annealing of the samples.  An argon ion etch was used to pattern the films into Hall bar pattern. Throughout the preparation we were careful to keep the temperature below 60$^\circ$C to avoid recrystallization of the indium oxide. After evaporating Ti-Au contact pads, the films were annealed in a 10 millitorr vacuum at about 55$^\circ$C for three weeks, during which time the room temperature sheet resistance decreased by about ten percent; a higher temperature anneal would have sped up the process but might have changed the microstructure of the film. Further details on the growth process are given in Ref.~\onlinecite{Steiner2005}

While InOx has been known as an amorphous low-carrier-density superconductor ($n\sim10^{20} - 10^{21}$ carriers/cm$^3$) and was used in many studies of SIT, different preparation methods result in different microstructure - hence different amount of ``disorder."  The reason disorder is put here in quotation marks is because of the complexity to quantify it when applied to the SIT. When films are granular, it is obvious that their SIT is dominated by Josephson tunneling among grains, and hence by phase fluctuations. However, even if films are inherently homogeneous, small and hardly detectable perturbations in the microstructure, may lead to large variations in the local strength of the superconducting order parameter and hence to effective granularity.  This effect is strongly magnified in the presence of a magnetic field (relevant to the H-SIT) which destroys weak links and thus enhance granularity. Pertaining to the InOx films used for the present study, we followed the process first described by by Kowal and Ovadyahu \cite{Kowal1984} showed that InOx can be made non-granular using a very low-temperature annealing technique. In their studies of similar films transmission electron micrographs were shown to be completely amorphous, and comparison with electron diffraction patterns from pure indium films ruled out the presence of In crystallites as small as $\sim 10\AA$ which were observed in films prepared by other methods. Nevertheless, the insulating side of the SIT was found to behave as a granular system.

\section{THEORETICAL}
\subsection{Duality and Self-Duality}

Consider a two-dimensional superconductor connected to a current source. Forcing a current density $\vec{J}$ through the superconductor means that a vortex will feel a Magnus (Lorentz) force
\begin{equation}
\vec{f} = \frac{1}{c}\vec{j} \times \vec{\Phi}_0=\frac{h}{2e}\vec{j}\times \hat{z}
\label{lorentz}
\end{equation}
where $\vec{\Phi}_0 =\hat z\Phi_0$ with $\Phi_0=hc/2e$. In the absence of vortices the applied current transforms inside the superconductor into supercurrent of Cooper-pairs (with charge 2e) with no dissipation. However, in the presence of vortices the phase changes by $2\pi$ each time a vortex crosses any imaginary line in the sample (e.g. the sample edge) resulting in a voltage determined by the Josephson relation 
\be
V = \frac{\hbar}{2e} \dot \phi= \frac{\Phi_0}{c} q_vn_vv_vL
\ee
where  $\dot\phi$ is the rate of chance of the phase, $q_v=\pm 1$ depending on the sense of the magnetic field (i.e. the sign of the vorticity), $n_v$ and $v_v$ are, respectively, the vortex density and mean velocity, and  $L$ is the size of the sample in the direction of the current. In other words, 
\begin{equation}
\vec{E} = -\frac{h}{(2e)^2}
\vec{j}^v \times \hat{z}
\label{charge_field}
\end{equation}
where $\vec{j}^v=2eq_vn_v\vec{v}_v$ is the vortex current density.  The minus sign in the equation reflects the fact that if the current is in the $+x$ direction, then according to equation~\ref{lorentz} the force on the vortex is in the $-y$ direction. In the same way, dividing Eqn.~\ref{lorentz} by the charge of a vortex and by the size of the sample in the direction of vortex propagation we obtain a ``vortex electric field"
\begin{equation}
\vec{E}^v= \frac{h}{(2e)^2}\vec{j}\times \hat{z}.
\label{vortex_field}
\end{equation}
Using the relations between current density and electric field
\be
j^v_a=\sigma^v_{ab}E^v_b,\ \ \  j_a=\sigma_{ab}E_b, \ \ \  E^v_a=\rho^v_{ab}j^v_b
\label{Ohms}
\ee
we arrive at the duality relation 
\be
-{\underline\epsilon} \ {\underline\sigma}\ {\underline\epsilon}=(4e^2/h)^2 \underline{\rho}^v,
\ee
where $\underline{\epsilon}=-\underline{\epsilon}^T$ with $\epsilon_{xx}=1$ is the Levi-Cevita tensor. In the case of an isotropic medium, this is equivalent to the duality relation given in Eq. 1 of the manuscript.

It is easy to see that if we invert the vortex resistivity tensor, the Cooper-pair and vortex conductivities are related through: 
\be
\underline{\sigma}=(4e^2/h)^2\frac{1}{(\sigma^v_{xx})^2+(\sigma^v_{xy})^2}[\underline{\sigma}^v]^T
\ee
where $[\underline{\sigma}^v]^T$ is the transpose of the vortex conductivity tensor. Now, the statement of \emph{Self Duality} is that (in our units) the magnitude of the current density of Cooper-pairs and vortices are equal, that is: $|\vec{j}|=|\vec{j}^v|$. Using Eqns.~\ref{charge_field}, \ref{vortex_field}, and \ref{Ohms} we can verify that self duality implies
\be
(\sigma^v_{xx})^2+(\sigma^v_{xy})^2=(4e^2/h)^2
\ee
and therefore
\be
\underline{\sigma}^T=\underline{\sigma}^v.
\ee
We note that these relations do not fully determine $\sigma_{xx}$ and $\sigma_{xy}$ independently. However, we can obtain an independent constraint from the following argument. The H-SIT is controlled at zero temperature by the magnetic field that is tuned through the critical point at $H_c$. Assuming that the conductivity tensor at the critical point is universal,  it cannot depend on $H_c$. This  implies that the Hall angle is zero, that is, $\vec{j}(H_c) \bot \vec{j}^v(H_c)$, and therefore $\sigma_{xy}(H_c)=0$. The consequences of this assertion are therefore that at criticality $\sigma_{xx}(H_c)=4e^2/h$, $\sigma_{xy}(H_c)=0$, and $\rho_{xy}(H_c)=0$ as stated in the main manuscript.

\subsection{Limiting behavior of $\rho^v_{xy}$ and $\sigma^v_{xy}$}
Starting from the insulating side, since $\rho_{xx}(T) \to \infty$ as $T \to 0$, it implies that $\sigma_{xx} = -\rho_{xx}/(\rho_{xx}^2 + \rho_{xy}^2) \to 1/\rho_{xx} \to 0$ as $T \to 0$. Therefore, in that regime, 
\be
\rho_{xy}^v =  \frac{-\sigma_{xy}^v}{(\sigma_{xx}^v)^2 + (\sigma_{xy}^v)^2} \rightarrow -(\rho_{xx}^v)^2\rho_{xy},
\ee
where the last equality used the fact that $\rho_{xy}=\sigma_{xy}^v$. For a finite $\rho_{xy}$ we obtain the general relation that on the insulating side $\rho_{xy}^v \propto (\rho_{xx}^v)^2$. This relation is identical to the condition $\sigma_{xy} \propto \sigma_{xx}^2$, found in Ref.~\onlinecite{Kivelson1992} for the ``Hall insulator" phase for which the longitudinal resistivity diverges, while the Hall resistivity approaches a constant as $T \to 0$.

While the term ``Hall insulator" was first coined for the insulating phase above the QHIT \cite{Kivelson1992}, for the present case of SIT it may be more revealing to analyze the superconducting side of the transition. The 
Cooper pairs conductivity diverges `$\sigma_{xx}(T)\to \infty$' and vortices become pinned, hence contributing a diverging vortex resistivity $\rho_{xx}^v(T) \to \infty$ as $T \to 0$. This implies that the vortex Hall conductivity is 
\be
\sigma_{xy}^v=\frac{-\rho_{xy}^v}{(\rho_{xx}^v)^2 + (\rho_{xx}^v)^2} \rightarrow \frac{-\rho_{xy}^v}{(\rho_{xx}^v)^2 } = -(\sigma_{xx}^v)^2\rho_{xy}^v
\label{vin1}
\ee
Using the fact that $\sigma_{xy}=\rho_{xy}^v$, which we observed to be finite on the superconducting side, we find that the dual condition for the Hall insulator for $H < H_c$ is $\sigma_{xy}^v \propto (\sigma_{xx}^v)^2$. This is equivalent to $\rho_{xy} \propto (\rho_{xx})^2$, a relation that was previously obtained by Vinokur {\it et al.} \cite{Vinokur1993} for the quenching  of vortex motion in disordered superconductors. On lowering the temperature pinning becomes relevant, and $\rho_{xx}^v$ displays thermally activated behavior, causing the measured $\rho_{xx}$ to decrease exponentially with temperature. In this regime the temperature dependence of the measured $\rho_{xy}$ is dominated by that of the measured $\rho_{xx}$, yielding Eqn.~\ref{vin1}.

\subsection{Composite Bosons and SIT}

Composite bosons in the quantum Hall effect are composed of an electron bound to an odd integer, $k$, of quanta of ``statistical'' flux, a construction that has a  precise meaning in terms of a Cherns-Simon field theory\cite{Zhang1989}.  The composite bosons are minimally coupled to an effective gauge field which is the sum of the electromagnetic gauge field, $A$, and the statistical gauge fields, $a$, where $a$ is a fluctuating (quantum dynamical) field.  However, to the extent that the fluctuations of $a$ about its mean-field (saddle-point) value can be treated as ``small,'' the response of the composite bosons can be treated in linear response.  In this case, the physical conductivity tensor (in units in which $e^2/h=1$) can be expressed\cite{Kivelson1992} in terms of the composite boson conductivity tensor, $\sigma_{ab}^{(cb)}$, according to the relations
\be
\sigma_{xx} = \frac{\sigma_{xx}^{(cb)}}{D^{(cb)}}\ \ \sigma_{xy} = k\left\{1- \frac{k[\sigma_{xy}^{(cb)}+k]}{D^{(cb)}}\right\}
\ee
\be 
D^{(cb)}=[\sigma_{xy}^{(cb)}+k]^2+[\sigma_{xx}^{(cb)}]^2.
\ee

The implications of this for the QHIT can be  illustrated by evaluating it in important limiting conditions:
\begin{itemize}
\item
The superconducting phase of the composite bosons in which $\sigma_{xx}^{(cb)} \to \infty$ corresponds to the quantum Hall phase with $\sigma_{xx} \to 0$ and $\sigma_{xy} \to k $.  
\item
The insulating phase of the composite bosons in which $\sigma_{xx}^{(cb)}\to 0$ and $\sigma_{xy}^{(cb)}\to 0$ corresponds to the insulating phase of the electrons in which $\sigma_{xx}\to 0$ and $\sigma_{xy}\to 0$.  
\item

Assuming the by now familiar universal values for the composite boson conductivity tensor at criticality, $\sigma_{xx}^{(cb)}=1$ and $\sigma_{xy}^{(cb)}=0$ (which was conjectured to hold at the QHIT\cite{Kivelson1992}) one finds $\sigma_{xx}=1/[1+k^2]$ and $\sigma_{xy}=k/[1+k^2]$ and correspondingly $\rho_{xx}= 1$ and $\rho_{xy} =k$.  Note that this last equality implies that the Hall resistance at criticality is equal to its value in the quantum Hall liquid phase -- this is highly suggestive that even in the limit $T\to 0$, the Hall resistance remains a continuous function of $H$ across the transition and into the proximate insulating phase, implying that it must be a Hall insulator.
\item
 As in the case of the SIT, a more careful analysis of the way in which the $\sigma_{ab}^{(cb)}$ vanishes  as $T\to 0$ is necessary to determine the character of the resistivity tensor in the insulating phase proximate to the QHIT.  Specifically,  exploiting the appropriate particle-vortex duality for the quantum Hall context, it was argued in Ref.~\onlinecite{Kivelson1992}  that $\sigma_{xy}^{(cb)} \sim [\sigma_{xx}^{(cb)}]^2$, as $T\to 0$, in which case $\rho_{xx} \to \infty$ but $\rho_{xy} \to k^3[1+k\alpha]/[1+k^2 + 2k^3\alpha + k^2\alpha^2]$ where $\alpha\equiv \lim_{T\to 0} \sigma_{xy}^{(cb)}/\sigma_{xx}^{(cb)}$.
\end{itemize}

While much of this discussion appears in Kivelson {\it et al.}, a more pedagogic review of these expressions, including  generalizations to transitions involving more complex quantum Hall liquid phases, is contained in Ref.~\onlinecite{Karlhede1993}.


\begin{thebibliography}{99}
\bibitem{Rhoc}
Uncertainty of order $\sim 10\%$ (or larger) is typical when determining the geometrical aspect ratio needed to calculate the 2D resistivity of disordered films.
\bibitem{maybe}
While a dangerously irrelevant operator could account for an additional temperature scale, smaller than the one governing $\rho_{xx}$, below which $\rho_{xy}$ would diverge, it cannot account for the sharp change in the $T$ dependence of $\rho_{xy}$ that is seen at $H_c^*$.
\end{thebibliography}
\end{document}